\documentclass[letterpaper,amsmath,
  amssymb,aps,prd,nofootinbib,
singlecolumn,
superscriptaddress,altaffilsymbol]{revtex4}

\usepackage{amsmath}
\usepackage{amsfonts}
\usepackage{amssymb}
\usepackage{graphicx}
\usepackage{psfrag}
\usepackage[hang]{footmisc}
\usepackage{setspace}
\newcommand{\beq}{\begin{equation}}
\newcommand{\eeq}{\end{equation}}
\newcommand{\bea}{\begin{eqnarray}}
\newcommand{\eea}{\end{eqnarray}}
\newcommand{\beqn}{\begin{equation*}}
\newcommand{\eeqn}{\end{equation*}}
\newcommand{\bean}{\begin{eqnarray*}}
\newcommand{\eean}{\end{eqnarray*}}

\newcommand*{\cref}[1]{Chapter~\ref{#1}}

\begin{document}
\title{Mixed constraints to inflationary models}

\author{Gabriel Germ\'an}
\email{Corresponding author: gabriel@icf.unam.mx}
\affiliation{Instituto de Ciencias F\'{\i}sicas, 
Universidad Nacional
Aut\'onoma de M\'exico, Cuernavaca, Morelos, 62210, Mexico}
\author{Juan Carlos Hidalgo}
\affiliation{Instituto de Ciencias F\'{\i}sicas, Universidad Nacional Aut\'onoma de M\'exico, Cuernavaca, Morelos, 62210, Mexico}
\author{Ariadna Montiel}
\affiliation{Instituto de Ciencias F\'{\i}sicas, Universidad Nacional
Aut\'onoma de M\'exico, Cuernavaca, Morelos, 62210, Mexico}

\date{\today}
\begin{abstract}
We show how to constrain inflationary models and reheating by using mixed constraints.  In particular we study the physics of the reheating phase after inflation from observational constraints to the inflationary stage. We show that it is possible to determine $\omega$, the equation of state during reheating, by using the reported values of the spectral index and the {\it full} number of $e$-folds $N(n_s,\omega)= N_H(n_s)+N_{re}(n_s,w)\approx 60$, which includes the accelerated expansion and the reheating phase. 
We show that the reheating number of $e$-folds $N_{re}$ is quite sensitive to this equation of state. Requiring $N_{re}>0$ and a sensible value for the thermalization scale $T_{re}$, demands in general a reheating phase with $\omega\neq 0$. We exemplify the constraints with two particular examples: We show how the Starobinsky model allows only large values of $T_{re}$ if the reheating phase is dominated by dust ($w =0$), and if Primordial Black Hole production is subdominant. For the case of $N=1$ Supergravity inflation, the extra parameter of the potential provides the necessary freedom to afford lower-scale thermalization in a dust-like reheating phase and yet our method serves to determine the rest of the observable parameters. 
\end{abstract}
\maketitle
\section{Introduction}

 Primordial Inflation represents the most successful paradigm to describe the physics of the early universe, previous to the hot big bang stage (seminal works are  \cite{Guth:1980zm,Linde:1984ir}, while useful reviews can be found in  \cite{Lyth:1998xn,Martin:2018ycu}). A primordial accelerated expansion of spacetime provides a coherent description of observations (e.g. the flatness of the universe and the homogeneity of CMB beyond the last scattering horizon scale), hard to explain from the big-bang model alone. Moreover, the most striking feature of inflation lies in the fact that quantum fluctuations of the inflaton field are able to provide the correct amplitude for the primordial density perturbations, which later evolve into the observed inhomogeneities of the subsequent stages in the Universe.   
    
Several models for the inflationary potential are able to accommodate the $60$ $e$-folds of expansion required to solve the flatness problem and produce an almost scale-invariant powerspectrum of the primordial perturbations (see Refs. \cite{Starobinsky:1980te,Mukhanov:1981xt,Starobinsky:1983zz,Whitt:1984pd,German:2004vf} for relevant examples). All of them provide an interpretation of the observed cosmic microwave background radiation (CMB) anisotropies \cite{Hinshaw:1996ut,Komatsu:2010fb,Aghanim:2018eyx,Akrami:2018odb}, and the large scale structure \cite{Percival:2001hw,Cole:2005sx,Benjamin:2007ys,Kilbinger:2014cea}. Most of this success is achieved by taking the evolution subject to a \textit{slow-roll} regime which relates the observable parameters to characteristics of the inflationary potential, as long as the inflaton evolves slowly down its potential \cite{Liddle:1994dx}. The challenge for advancing probes of the inflationary universe is to provide data in order to constrain and discriminate between models of inflation through parameters like the deviation from scale-invariance of the powerspectrum, i.e. the spectral index $n_s$, and also its scale dependence encoded in the running $n_{sk}$, as well as the detection of tensorial modes from the primordial perturbations (most recent surveys are \cite{Aghanim:2018eyx,Akrami:2018odb,Ade:2018gkx} while future missions are already in preparation \cite{Delabrouille:2017rct}). 
    
A crucial aspect of the inflationary theory is the transition to the radiation-dominated era of the standard Big-Bang theory. The process of energy transfer from the inflaton field to the relativistic species is generically named reheating \cite{Kofman:1997yn,Greene:1997fu,Allahverdi:2010xz,Amin:2014eta}. 
The elusive observables from this period result in a poor understanding of the physics at the end of the inflationary era. On the observational side, signatures from reheating could be encoded in the high frequency end of the gravitational wave spectrum \cite{Antusch:2016con,Amin:2018xfe,Liu:2017hua}, and in the abundance of Primordial Black Holes (PBHs) and their evaporation products \cite{Jedamzik:2010dq,Hidalgo:2011fj,Torres-Lomas:2014bua,Hidalgo:2017dfp,Carr:2018nkm}. On the theoretical side, it is known that the fluid approximation can describe the phase after accelerated expansion if the equation of state is $\omega> -1/3$. The coherent oscillations of the inflaton at the bottom of its potential are accounted by an effective $\omega = 0$. If there are resonant oscillations with an auxiliary field, these lead to instabilities that rapidly excite the coupled relativistic species resulting in a rapid preheating phase, after which the thermalization process may have a range of values $0 < \omega < 0.25$ \cite{Podolsky:2005bw}. Thus models of reheating disfavour $\omega \gtrsim 0.3$ and strictly require $\omega>-1/3$. In an important reference to this work, it has recently been suggested \cite{Dai:2014jja} that the restrictions to the spectral index $n_s$ and the dependence of $N_{re}$ (the  number of $e$-foldings elapsed during reheating), on the constrained parameters of inflation can provide a complementary constrain to the parameters of a power-law inflationary potential.

 In this paper, we take an alternative view on the formula for $N_{re}$ (presented below in Eq.~\eqref{NRE}). We provide constraints to the equation of state of reheating by evaluating the \textit{full} number of $e$-foldings  $N_{H}+N_{re}$ and by considering the theoretical limitations for this quantity. We exemplify our technique with two inflationary models, namely the Starobinsky model \cite{Starobinsky:1980te,Mukhanov:1981xt,Starobinsky:1983zz} and a working version of inflation inspired in $N=1$ Supergravity \cite{Ellis:1982dg,NANOPOULOS198341,OVRUT1983161,HOLMAN1984343,Ross:1995dq,German:2004vf}.  The results show that a certain degree of fine-tunning is required if one is to impose a dust-like $\omega$ for the reheating phase. Moreover, avoiding overproduction of PBHs during reheating yields extra constraints that further bound $\omega$ and the temperature at which the universe thermalises before the radiation era. 

\section{Reheating after slow-roll inflation }\label{Sec:II}

In a very interesting paper \cite{Dai:2014jja} the possibility of constraining inflationary models from reheating was discussed (see also \cite{Munoz:2014eqa} for the method applied to specific models of inflation). By combining a constraint on the total amount of expansion \cite{Liddle:2003as}  with the post-inflationary expansion of the universe, accounting for the evolution of energy density and temperature, it was shown that the number of $e$-folds during reheating $N_{re}\left(\omega, n_s \right)$ and the reheat temperature $T_{re}\left(\omega, n_s\right)$ can be expressed as functions of the equation of state parameter $\omega$ and the spectral index $n_s$, as well as model-dependent parameters. The number of $e$-folds during reheating and the thermalization temperature $T_{re}$ can be treated as functions of the spectral index for various values of $\omega$ and of a parameter characteristic of the specific model in question. Given that the possible  values of the spectral index reported by the Planck satellite are now a narrow range \cite{Akrami:2018odb}, we intend to use inflationary conditions to constrain the physics of the reheating phase 

With this in mind we now display the two relevant formulas obtained in \cite{Dai:2014jja}.
The number of $e$-folds during reheating $N_{re}\left(\omega, n_s \right)$  is given by 
\beq
\label{NRE}
N_{re}\left(\omega, n_s \right)= \frac{4}{1-3\, \omega}\left( -N_H-\frac{1}{4} \ln[\frac{\rho_e}{H_H^4} ]-\ln[\frac{k}{a_0 T_0}] -\frac{1}{4} \ln[\frac{30}{\pi^2 g_{re} } ]  -\frac{1}{3} \ln[\frac{11 g_{s,re}}{43} ]     \right),
\eeq
where the subindex $_H$ in any quantity designates the value of that quantity when perturbations are produced, some $50$ $-$ $60$ $e$-folds before the end of inflation,
the latter denoted by the subindex $_e$. Thus, $N_H$ is the number of $e$-folds during inflation from $\phi_H$ to $\phi_e$. In the formula above {the scales $H_H$, as well as $\rho_e$ and $N_H$} can also be functions of model-dependent parameters if present. The quantities $g_{re}$ and $g_{s,re}$ represent the effective number of relativistic species upon thermalization and of light species for entropy at reheating, respectively.

To evaluate the previous expression as a function of $\omega$, $n_s$ and of the parameters of the model (if any) we only need to know $\phi_H=\phi_H(n_s)$ and  $\phi_e$. The latter is usually determined by the time when slow-roll conditions are violated, another model-dependent value that we shall exemplify below. 

For future reference we introduce here some useful formulas. In the slow-roll approximation, the spectral indices are given in terms of the slow-roll parameters of the model, these are
 (see e.g. \cite{Liddle:1994dx}, \cite{Lyth:1998xn})
\begin{equation}
\epsilon \equiv \frac{M^{2}}{2}\left( \frac{V^{\prime }}{V }\right) ^{2},\quad
\eta \equiv M^{2}\frac{V^{\prime \prime }}{V}, \quad
\xi_2 \equiv M^{4}\frac{V^{\prime }V^{\prime \prime \prime }}{V^{2}}.
\label{Slowparameters}
\end{equation}%
Primes denote derivatives with respect to the inflaton field $\phi$ and $M$ is the
reduced Planck mass $M\equiv \left(8 \pi G\right)^{-1/2} = 2.44\times 10^{18} \,\mathrm{GeV}$, and which we set
$M=1$ in most of what follows. 
In the slow-roll approximation, observables are given by (see e.g. \cite{Liddle:1994dx}, \cite{Lyth:1998xn}, \cite{Planck:2013jfk,Ade:2015lrj})
\begin{eqnarray}
n_{t} &=&-2\epsilon =-\frac{r}{8} , \label{nt} \\
n_{s} &=&1+2\eta -6\epsilon ,  \label{ns} \\
n_{sk} &=&\frac{d n_{s}}{d \ln k}=16\epsilon \eta -24\epsilon ^{2}-2\xi_2, \label{nsk} \\
A_s(k) &=&\frac{1}{24\pi ^{2}} \frac{\Lambda^4}{%
\epsilon _H}, 
\label{A} 
\end{eqnarray}
where $n_{sk}$ is the running of the scalar index. The scalar powerspectrum amplitude at wave number $k$ is $A_s(k)$ and the scale of 
inflation is $\Lambda$, with $\Lambda \equiv V_{H}^{1/4}$. The values of the above parameters  inferred from cosmological surveys are $n_s = 0.9649 \pm 0.0042$ and $n_{sk}=-0.0045 \pm 0.0067$ both at 68$\%$ confidence \cite{Akrami:2018odb}, also $A_s(k=0.05)=2.1955 \times 10^{-9}$ \cite{Planck:2013jfk}, and $r_{0.05}<0.07$ at 95$\%$ confidence \cite{Ade:2018gkx}.

A final quantity of physical relevance is the thermalization temperature at the end of the reheating phase,
\beq
\label{TRE}
T_{re}\left(\omega, n_s\right)=\left( \frac{30\, \rho_e}{\pi^2 g_{re}} \right)^{1/4}\, e^{-\frac{3}{4}(1+\omega)N_{re}\left(\omega\right)}\,.
\eeq

\noindent This is a function of the number of $e$-folds of reheating and the equation of state at that stage. An important constraint for $T_{re}$ comes directly from the overproduction of primordial black holes (PBHs) in the case of a soft equation of state during reheating. Indeed, if $\omega \approx 0$, the overdensities at the end of inflation have enough amplitude as to produce a large number of black holes with mass $M_{\rm PBH} \sim 10^{-18} M_{\odot}$ if $10^8 < T_{re} / \mathrm{GeV}< 10^9$ \cite{Hidalgo:2017dfp}. Such PBHs would be evaporating today due to Hawking radiation and the ammount of PBHs produced at this range of temperatures is enough to saturate the observational constraint $\beta \equiv \rho_{\rm PBH}/ \rho_{c} \lesssim 10^{-27}$ \cite{Carr:2009jm,Carr:2016hva}. We shall be observing this constrain to the reheating temperature in the results reported below.  
\section{The Starobinsky model} \label{The model} 
The potential of the Starobinsky model \cite{Starobinsky:1980te,Mukhanov:1981xt,Starobinsky:1983zz} is given by \cite{Whitt:1984pd}:
\beq
\label{staropot}
V= V_0 \left(1- e^{-\sqrt{\frac{2}{3}}\phi} \right)^2.
\eeq
From Eq.~(\ref {ns}) the solution for $\phi_H$ in terms of the spectral index is,
\beq
\label{fih}
\phi_H= \sqrt{\frac {3}{2}}\ln      \left(\frac{7-3n_s+4\sqrt{4-3n_s}}{3(1-n_s)}  \right) ,
\eeq
while the end of inflation is given by the solution to the equation $\epsilon =1$ at $\phi_e$:
\beq
\label{fie}
\phi_e= \sqrt{\frac {3}{2}}\ln\left(1+\frac{2}{\sqrt{3}}\right) .
\eeq
From Eqs.~(\ref{fih}) and (\ref{nt}) we find the tensor-to-scalar ratio $r$ as a functions of $n_s$,
\beq
\label{r}
r= \frac {4}{3} \left(5-3n_s-2\sqrt{4-3n_s} \right) .
\eeq
The number of $e$-folds during inflation from $\phi_H$ to $\phi_e$ is
\beq
\label{NH}
N_H= \frac {1}{4} \left( 3e^{\sqrt{\frac{2}{3}}\phi_H} -\sqrt{6} \,\phi_H \right)-\frac {1}{4} \left( 3e^{\sqrt{\frac{2}{3}}\phi_e} -\sqrt{6} \,\phi_e \right) .
\eeq
From the equation for the density perturbations (\ref{A}) at wave number $k$ the potential at $\phi_H$ is given by
\beq
\label{VH}
V_H= 3 H_{H}^2=\frac{3}{2}\pi^2 r A_s\,,
\eeq
while the potential at the end of inflation at $\phi_e$ can be related to $V_H$  by means of the formula
\beq
\label{Ve}
V_e= \frac{\left(1- e^{-\sqrt{\frac{2}{3}}\phi_e} \right)^2        }{\left(1- e^{-\sqrt{\frac{2}{3}}\phi_H} \right)^2    }V_H\,,
\eeq
from where it follows an expression for the energy density $\rho$ at $\phi_e$, that is,
\beq
\label{roe}
\rho_e= \frac{3}{2} V_e\,.
\eeq
\begin{figure}[tb]
\begin{center}
\includegraphics[width=8cm]{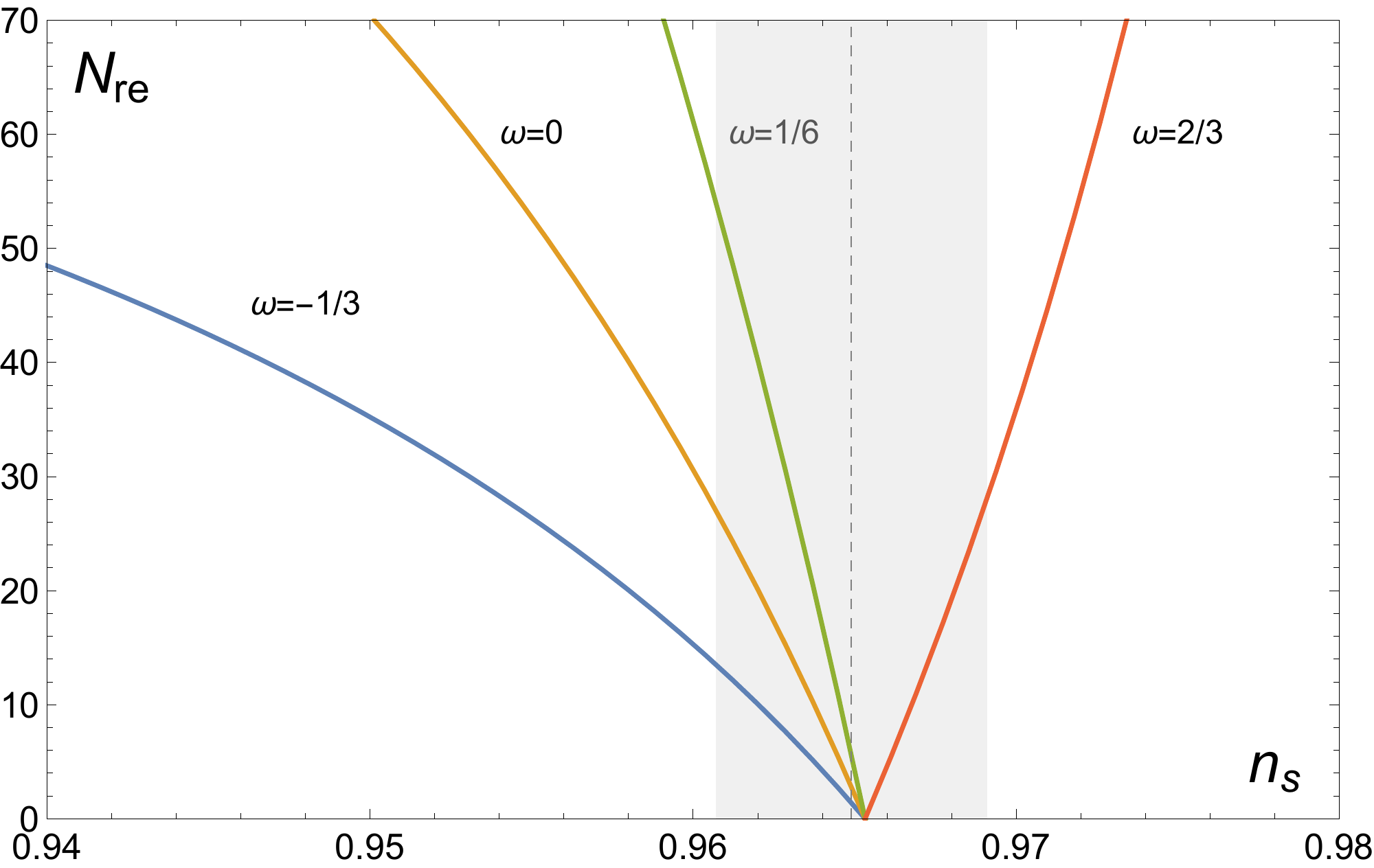}
\caption{\small Following \cite{Dai:2014jja} the plot shows the number of $e$-folds during reheating $N_{re}$ for the Starobinsky model as a function of the spectral index $n_s$, as given by Eq.~(\ref{NRE}), for various values of $\omega$. The shaded region covers the $1-\sigma$ values of $n_s$ according to Planck-2018 \cite{Akrami:2018odb}. (See in contrast, Fig.~\ref{Ndeomega} where the {\it full} number of $e$-folds $N=N_H+N_{re}$ is plotted against $\omega$ for various $n_s$.}
\label{NreStaro}
\end{center}
\end{figure}
\begin{figure}[tb]
\begin{center}
\includegraphics[width=8cm]{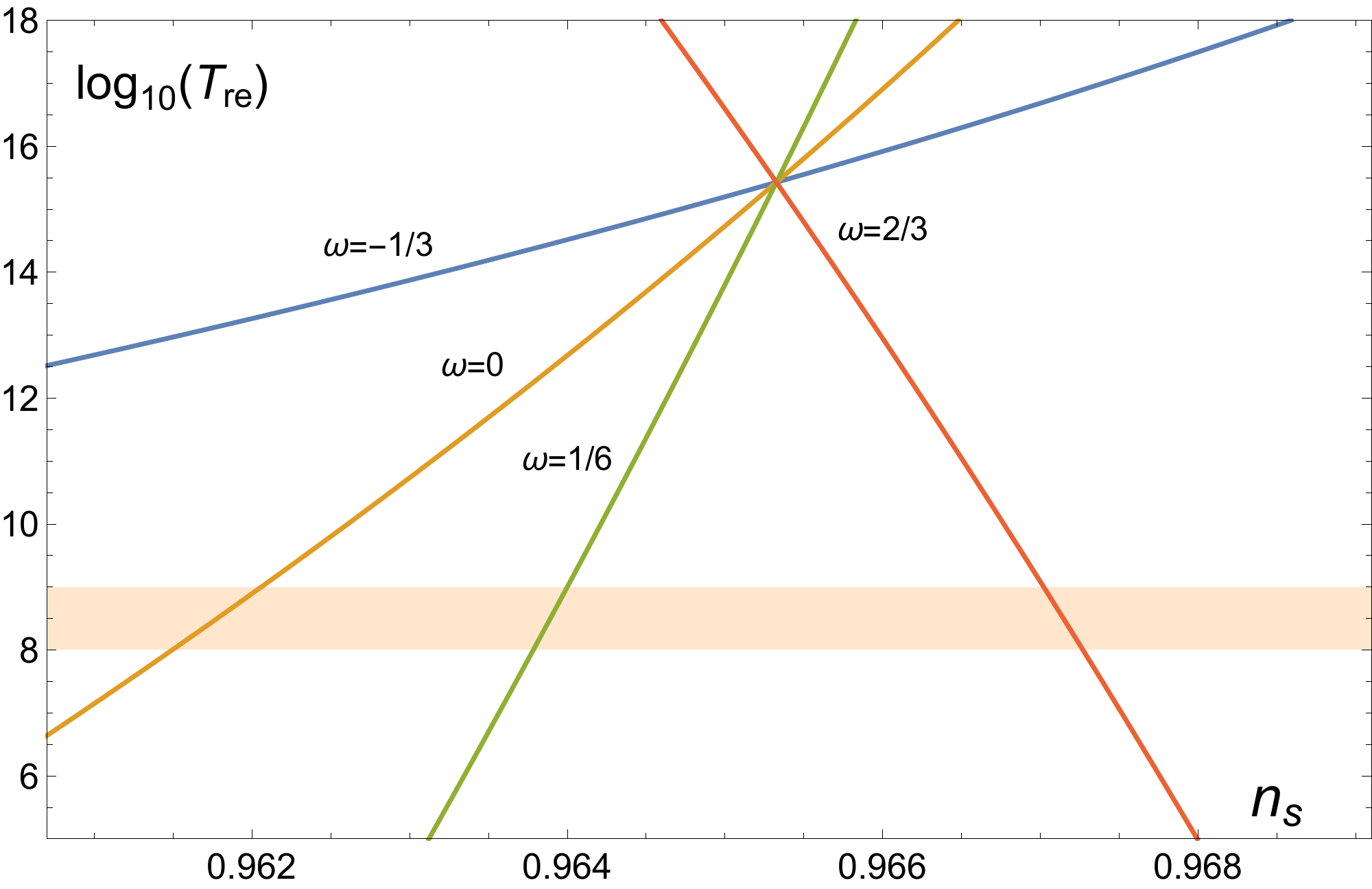}
\caption{\small Plot of the logarithm of the reheat temperature $T_{re}$ in the Starobinsky model as a function of the spectral index $n_s$ in the $1-\sigma$ range reported by Planck-2018, as given by Eq.~(\ref{TRE}). Some values of the equation of state during reheating $\omega$ are shown. The orange shaded region indicates the range of reheating temperature values where PBHs are overproduced if $\omega =0$ (see discussion at the end of Sec.~\ref{Sec:II}.}
\label{TreStaro}
\end{center}
\end{figure}
For the numerical calculations we follow \cite{Dai:2014jja} and take $g_{re} \approx g_{s,re} =100$, $\frac{k}{a_0 T_0}=1.36 \times 10^{-27}$.  A plot of $N_{re}$ as a function of $n_s$ for various values of $\omega$ is shown by Fig.~\ref{NreStaro} and the reheat temperature given by Fig.~\ref{TreStaro}. 
These figures have already been given in \cite{Cook:2015vqa} for the Starobinsky model (without the PBHs constraint). Here we would like to extend this work by plotting first both the {\it full} number of $e$-folds (meaning $N=N_H+N_{re}$) as a function of $\omega$  and the reheat temperature $T_{re}$ for a range of values of $n_s$ given by Planck 2018 \cite{Akrami:2018odb}. Then using (for definitiveness) the central value for the spectral index $n_s = 0.9649$ and imposing the condition that $N=60$ 
we extract the value of the equation of state parameter $\omega$ and calculate the number of $e$-folds during inflation $N_H$ as well as during reheating $N_{re}$. The reheat temperature $T_{re}$ is also determined together with the usual observables such as the running and the tensor-to-scalar ratio. 
\begin{figure}[tb]
\begin{center}
\includegraphics[width=8cm]{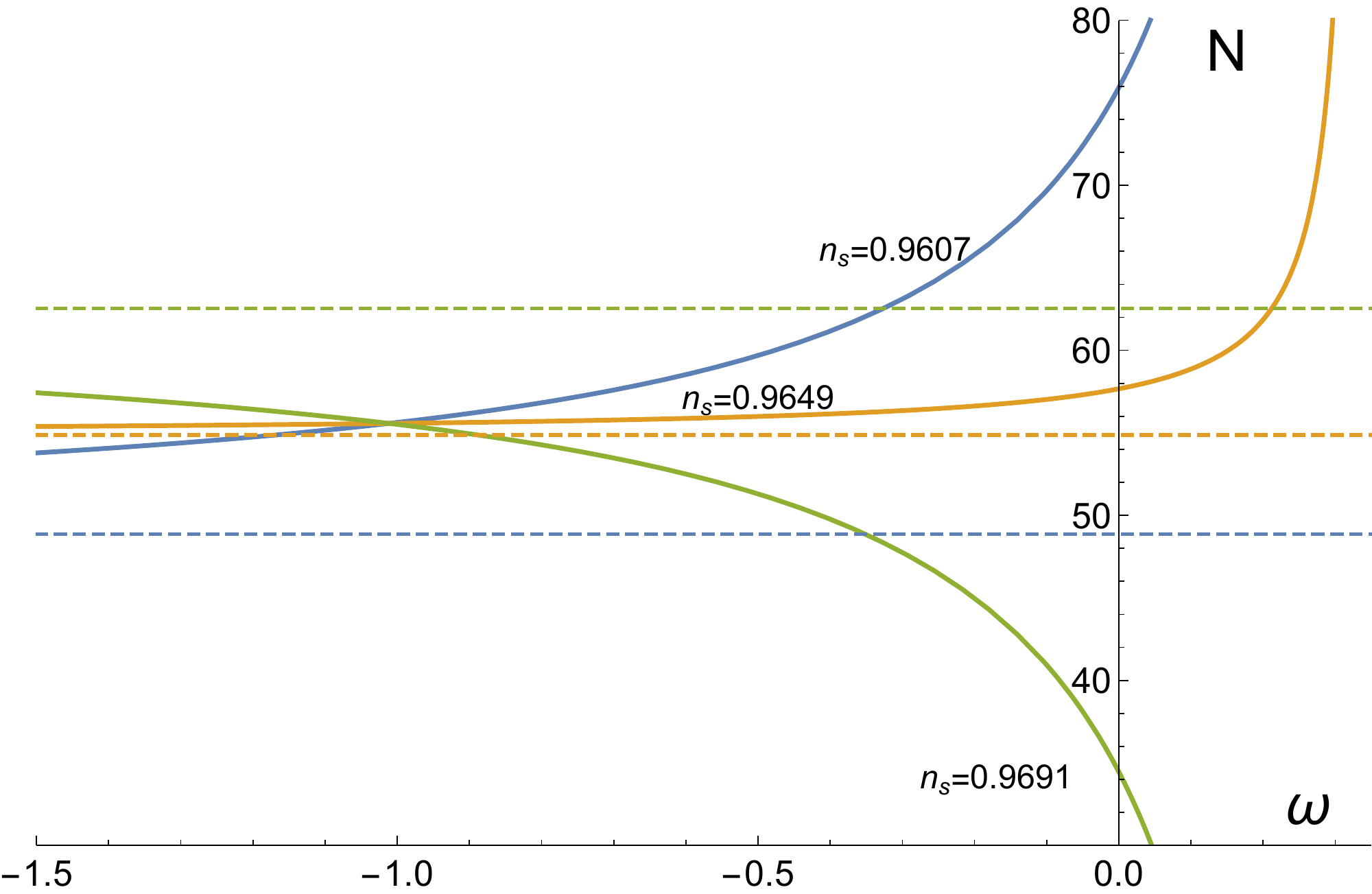}
\caption{\small Plot of $N\equiv N_H+N_{re}$  against $\omega$ for values of the spectral index $n_s = 0.9649 \pm 00042$ as reported by Planck-2018. Dashed horizontal lines show $N_H$ for each case. {Since $N$ becomes a decreasing function for $n_s\gtrsim 0.9653$, the number of $e$-folds elapsed during reheating $N_{re}$ becomes negative. This is illustrated with the set of green lines for the case $n_s = 0.9691$}}
\label{Ndeomega}
\end{center}
\end{figure}
\begin{figure}[tb]
\begin{center}
\includegraphics[width=8cm]{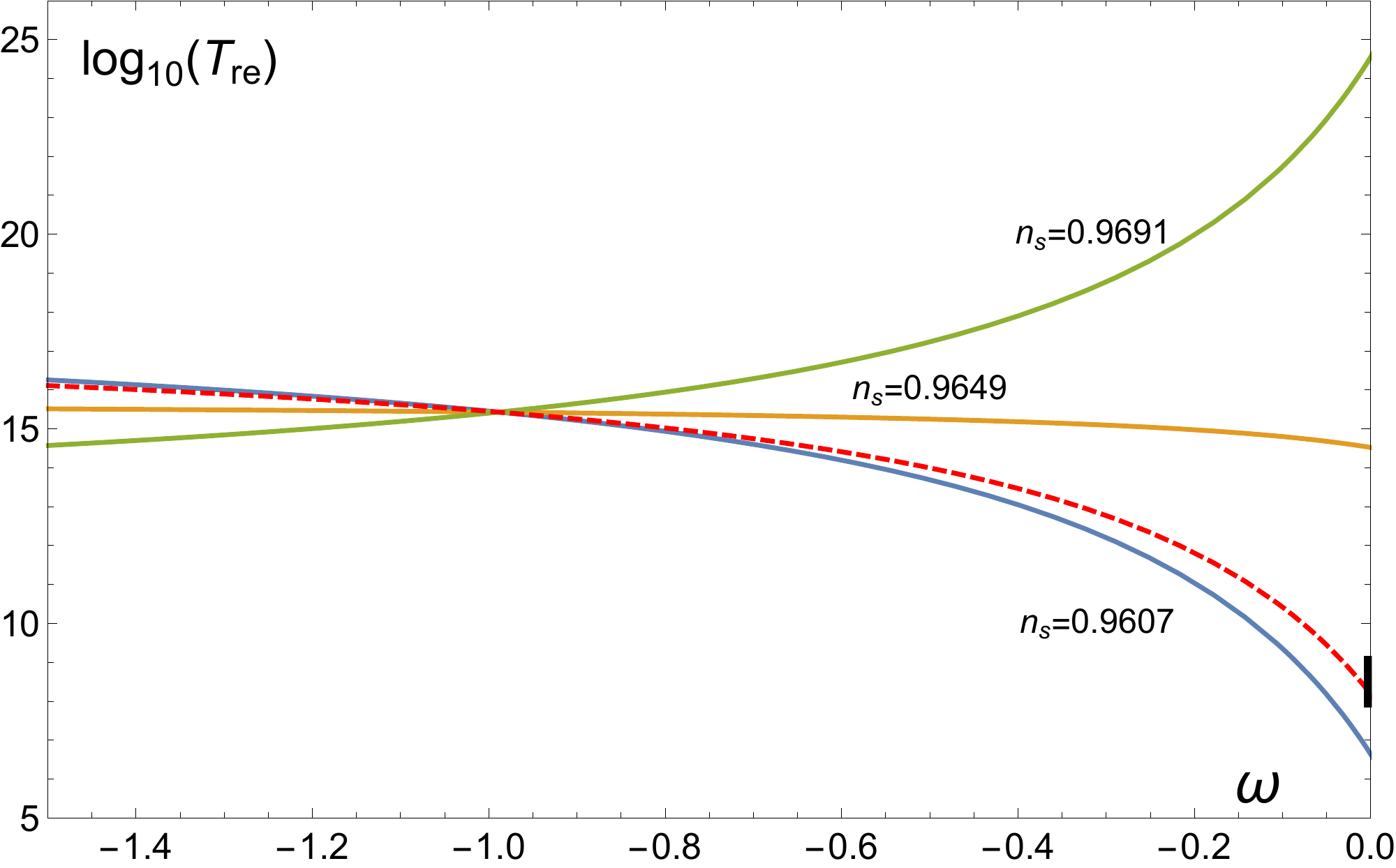}
\caption{\small Plot of $T_{re}\left(\omega\right)$ for the three values of the spectral index described in Fig.~\ref{TreStaro} plus $n_s = 0.9616$ (red dotted line). For the central value of the spectral index $n_s=0.9649$ and a {\it full} number of $e$-folds $N=N_H+N_{re} \approx 60$, $T_{re} \approx 3.3 \times 10^{13}$~GeV. Models intersecting the shaded region on the ordinate axis (e.g. $n_s = 0.9616$) will overproduce PBHs as discussed at the end of Sec.~\ref{Sec:II}.}
\label{Treheating}
\end{center}
\end{figure}

For the Starobinsky model the term between brackets in Eq.~(\ref{NRE}) becomes negative for $n_s \geq 0.9653$ for any $\omega<1/3$. Of course one can continue working with Eq.~(\ref{NRE})  if $\omega>1/3$ but one may wonder if this is physically acceptable. To illustrate this effect, Fig.~\ref{Ndeomega} shows the {\it full} number of $e$-folds $N=N_H+N_{re}$ as a function of $\omega$ for $n_s = 0.9607, 0.9649$ and $0.9691$ corresponding to the central value and the $1-\sigma$ values from Planck 2018 results. {Determining $N_H$ from Eq.~\eqref{NH} (plotted as a horizontal dotted line in each case), it is evident that the elapsed number of $e$-folds during reheating becomes negative if the sum is required to match the plotted \textit{full} $N$ for the case $n_s = 0.9691$}. Indeed, the curve with $n_s = 0.9691$ in Fig.~\ref{Ndeomega} makes no sense since it implies a {\it negative} $N_{re}$ for $\omega<1/3$. 

\begin{figure}[tb]
\begin{center}
\includegraphics[width=8cm]{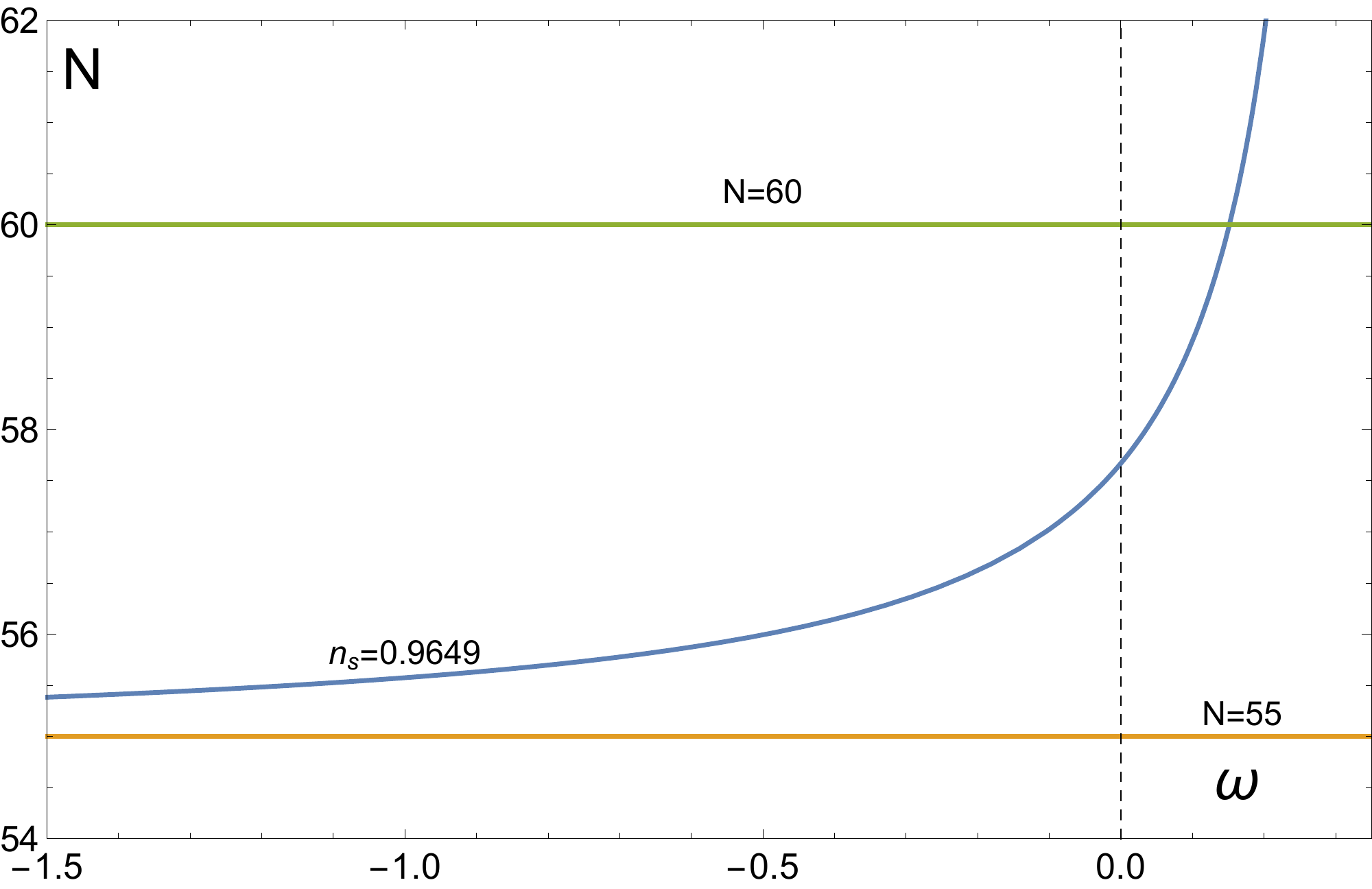}
\caption{\small Plot of the \textit{full} number of $e$-folds $ N \equiv N_H+N_{re}$ as a function of $\omega$ for the Planck-2018 mean value $n_s=0.9649$ 
The bottom horizontal line marks $N = 54.88$ $e$-folds elapsed due to the inflationary epoch and the top line marks the full 60 $e$-folds. A full number of $e$-folds $N=60$ corresponds to $\omega \approx 0.1479$. Thus, $N_{re} = 5.12$ $e$-folds elapsed during the reheating epoch.}
\label{N60}
\end{center}
\end{figure}

Similarly, Fig.~\ref{Treheating} shows the reheat temperature $T_{re}$ as a function of $\omega$ for the same values of the spectral index. The increasing curve is again inconsistent since according to Eq.~(\ref{TRE}) $T_{re}$ should be a decreasing function of $\omega$. Note also that a few values of $n_s \approx 0.9616$ will be restricted to equations of state $\omega \neq 0$. This is because the range of temperature values at thermalization would overproduce PBHs as discussed at the end of Sec.~\ref{Sec:II}. Conversely, one can take the central value of the spectral index according to Planck 2018 $n_s=0.9649$ and a number of $e$-folds (coming from the inflationary plus the reheating epochs) equal to 60. This case is plotted in Fig.~\ref{N60}, from where it follows that the equation of state parameter during reheating takes the value $\omega_{re}\approx 0.1479$. On the other hand fixing $\omega=0$ as in the canonical reheating scenario gives a total of $N\approx 58$ $e$-folds of inflation and some other plausible results for $T_{re}$ and the other observables. Table~\ref{table1} provides numerical results stressing this point.
\begin{table}[h!]
  \centering
  \begin{tabular}{cccccccc}
    \\
$n_s (\mathrm{fixed})$  &\,\, $N$  &\,\, $\omega$  &\,\, $N_H$ & \,\,$N_{re}$ & \,\,$r$ & \,\,$n_{sk}$ & \,\, $T_{re}\,(GeV)$
\\
\\
0.9649  & \quad fixed to 60& \quad $0.1479$  & \quad $54.88$ & \quad $5.12$ & \quad $3.5 \times 10^{-3}$ & \quad $-6.2 \times 10^{-4}$ & \quad $3.3 \times 10^{13}$\\
\\
$0.9649$  & \quad $N=57.67$  & \quad  fixed to 0  & \quad $54.88$ & \quad $2.80$ & \quad $3.5 \times 10^{-3}$ & \quad $-6.2 \times 10^{-4}$ & \quad $2.4 \times 10^{16}$\\
    \\
\end{tabular}
\caption{In the first row the scalar spectral index $n_s$ and the {\it full} number of $e$-folds $N\equiv N_H+N_{re}$ are fixed. This sets $\omega$ to the value 0.1479, from here all other quantities are fixed. The second row fixes $\omega=0$ and the spectral index. In this case, the {\it full} number of $e$-folds becomes $N=57.67$ and all other quantities get fixed values. Using constraints from inflation give an exotic reheating phase while mixing the constraints with $n_s=0.9649$ on the one hand, and the canonical reheating scenario (where $\omega=0$) on the other, result in a sufficient number of $e$-folds and a reheating temperature characteristic of the GUT scale.}
 \label{table1}
\end{table}
\section{The Supergravity model} \label{Sugra}
\begin{figure}[tb]
\begin{center}
\includegraphics[width=8cm]{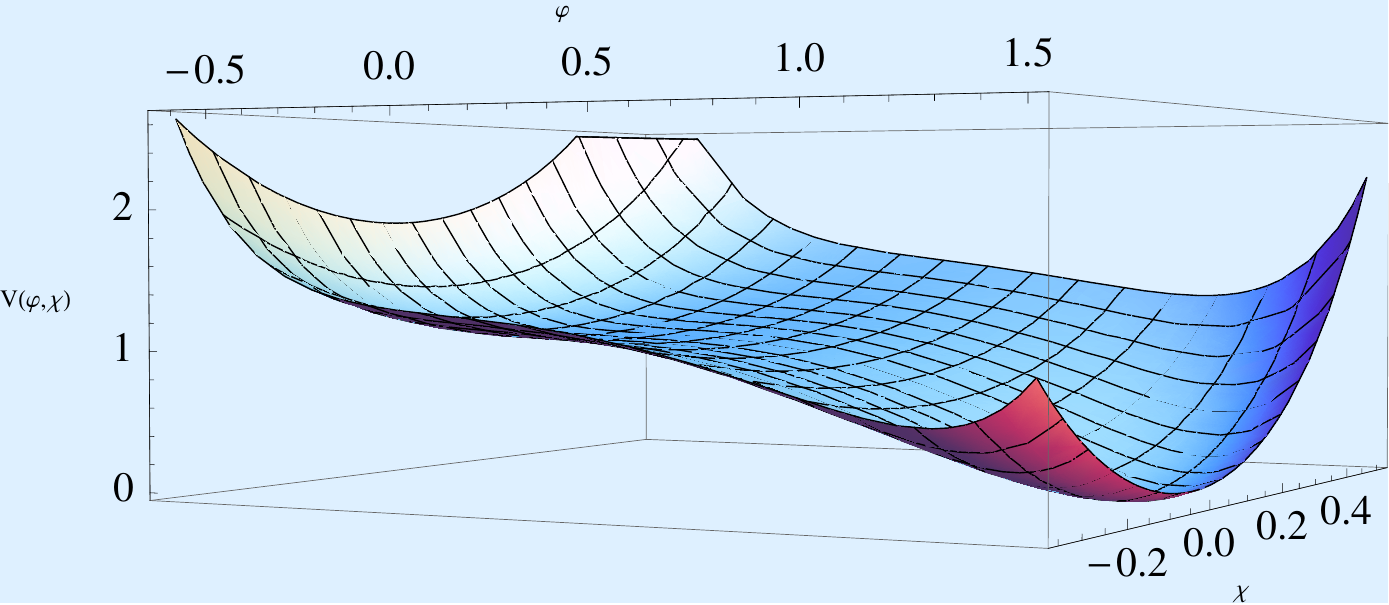}
\caption{\small Plot of the full potential given by Eqs.~(\ref{sugrapot3}) and (\ref{complexfield}) along the $\chi$ and  $\phi$ directions. The $\chi$ direction is stable and we can then study the inflationary $\phi$-direction safely on its own. The $\phi$ potential is given by Eq.~(\ref{sugrapot4}) and is illustrated in Fig.~\ref{potfi}.}
\label{Pot3D}
\end{center}
\end{figure}

\begin{figure}[tb]
\begin{center}
\includegraphics[width=8cm]{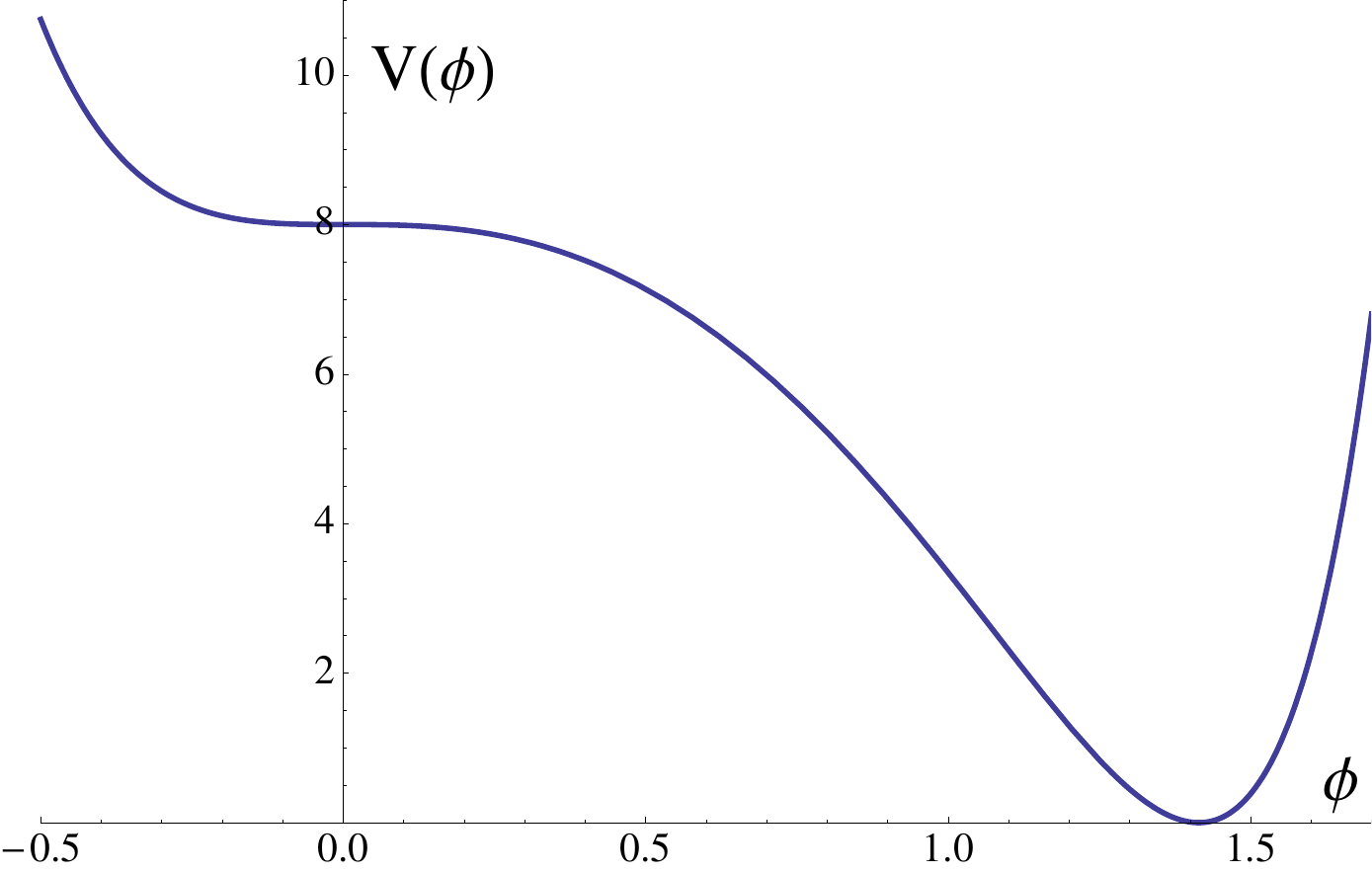}
\caption{\small Schematic plot of the inflationary potential  given by Eq.~(\ref{sugrapot4}).  The minimum occurs for $\phi = \phi_0$ and the potential is flat at the origin if $\phi_0 = \sqrt{2}$. However the slope of the potential at the origin should be fixed by using Planck data \cite{German2018}. An interesting feature of this model is that inflation can be understood as a transient phenomenon with inflation beginning and ending at the points $|\eta|=1$ with a finite {\it full} number of $e$-folds not much bigger than the $\approx 60$ requiered for solving cosmological problems \cite{German:2004vf}.}
\label{potfi}
\end{center}
\end{figure}

The second example to study comes from Supergravity. The piece of the $N=1$ Supergravity model we are presently interested in is given by the action  \cite{Cremmer:1982en}:
\beq
\label{sugraction}
I= -\int d^4 x \sqrt {-g}\left(\frac{1}{2}R + G_i^{\,\,j} \partial_{\mu}\Phi^i \partial_{\nu}\Phi^*_jg^{\mu\nu} + V\right),
\eeq
where the Kahler metric $G_i^{\,\,j}$ is defined by $G_i^{\,\,j}=\partial^2 G/\left(\partial \Phi^*_j\partial\Phi^i\right)$ and the 
Kahler function is $G( \Phi^i, \Phi^*_i)=K( \Phi^i, \Phi^*_i) + \ln |W(\Phi^i)|^2$, with $W$ a holomorphic function of $\Phi^i$ called superpotential and $K$ is the Kahler potential, a real function depending on the superfields $\Phi^i$ as well as their conjugates $ \Phi^*_i$. 
The \textit{F-term} of the scalar potential is given in terms of the real function $G$ as follows:
\beq
\label{sugrapot1}
V= e^G\left(G_i(G^{-1})^i_jG^j-3  \right).
\eeq
For the model of our present interest we only need a single chiral superfield $\Phi$ with scalar component $z$.
Thus, the potential is given by
\beq
\label{sugrapot2}
V= e^K\left(F_z (K_{z z^*})^{-1} F^*_{z^*}-3|W|^2  \right),
\eeq
where
\beq
\label{derivative}
F_z \equiv \frac{\partial W}{\partial z}+\frac{\partial K}{\partial z} W, \quad K_{zz^*}\equiv \frac{\partial^2 K}{\partial z\partial z^*} .
\eeq
To first approximation we take the Kahler potential to be of the canonical form
\beq
\label{kpot}
K(z,z^*)= z z^*+ \cdot  \cdot  \cdot \, ,
\eeq
with superpotential  \cite{Ellis:1982dg,NANOPOULOS198341,OVRUT1983161,HOLMAN1984343,Ross:1995dq}
\beq
\label{superpot}
W(z)= f(z_0)(z-z_0)^2, 
\eeq
where $f(z_0)$ is a constant with dimensions of mass which we simple take as $\Lambda$. The scalar potential becomes
\beq
\label{sugrapot3}
V=\Lambda^2 e^{z z^*}(z-z_0)(z^*-z^*_0) \left[ 4+z^*(z+z_0+z(z-z_0)z^*)+(z-3z_0-z^2z^*+zz^*z_0)z^*_0\right].
\eeq
Writing $z$ in terms of real components
\beq
\label{complexfield}
z=\frac{1}{\sqrt 2}(\phi+i\chi),
\eeq
we find that the $\chi$-direction is a stable direction of the full potential, Fig.~\ref{Pot3D}.
Thus we set $\chi=0$ and study the potential along the inflationary $\phi$-direction only \cite{German:2004vf}:
\beq
\label{sugrapot4}
V=\Lambda^2 e^{\frac{1}{2}\phi^2}(\phi-\phi_0)^2 \left[ 2+\frac{1}{8}(\phi-\phi_0)\left(6\phi_0+\phi(2+\phi^2-\phi\phi_0)\right)\right].
\eeq
This potential is illustrated in Fig. \ref{potfi} for some typical values of the parameters. This potential has a minimum at $\phi_0$ with vanishing energy. 
We redefine $\phi_0$ in terms of a new parameter $s$ as
\beq
\label{slope}
\phi_0=s/8+\sqrt {2},
\eeq
and calculate the derivative of $V$ at the origin
\beq
\label{firstderivativeV}
V^{\prime}(\phi=0)/\Lambda^2= s+\frac{3s^2}{16\sqrt{2}}+\frac{s^3}{256}.
\eeq
\begin{figure}[tb]
\begin{center}
\includegraphics[width=8cm]{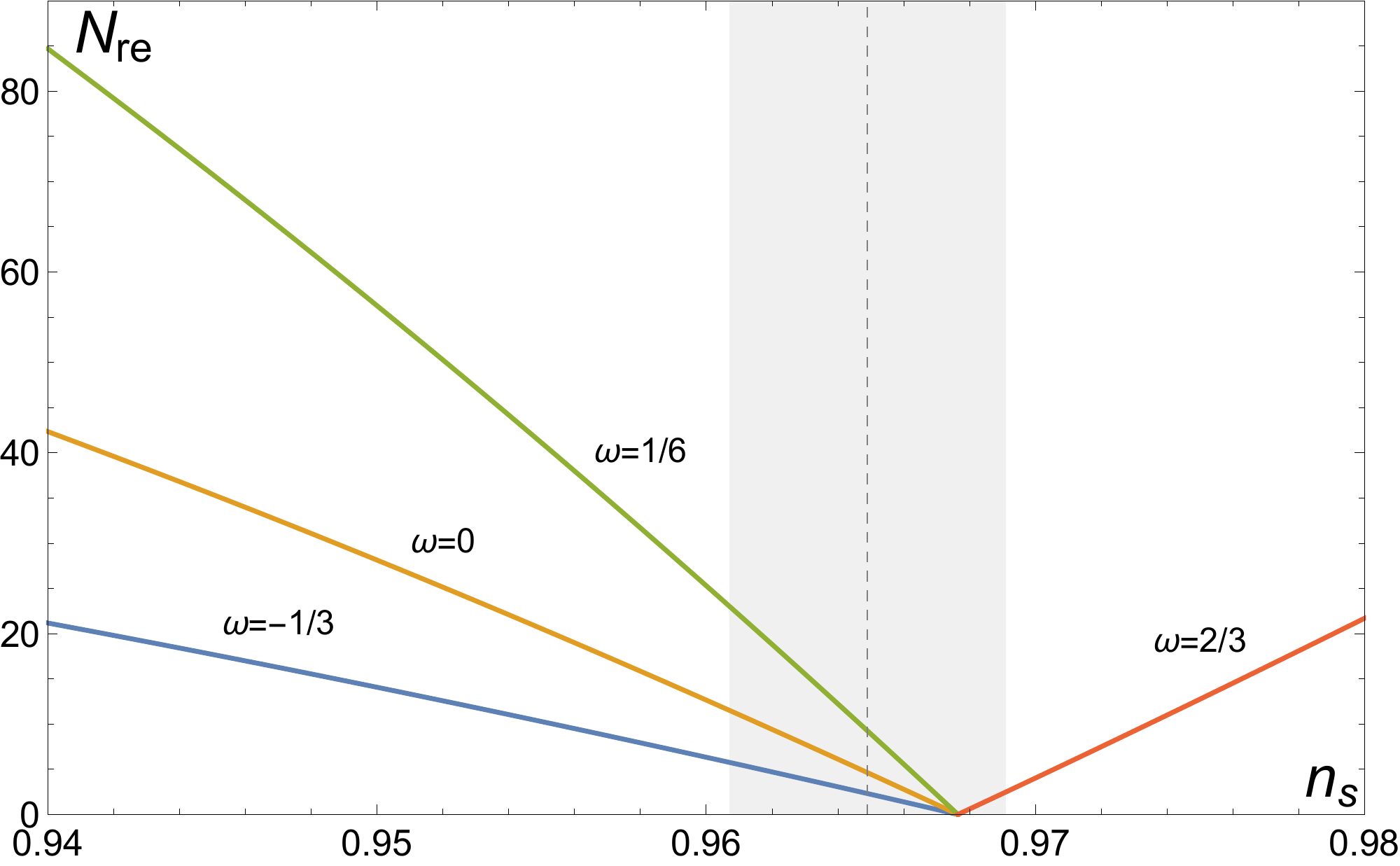}
\caption{\small Following \cite{Dai:2014jja} we plot the number of $e$-folds during reheating $N_{re}$ as a function of the spectral index $n_s$ as given by Eq.~(\ref{NRE}) for various values of the equation of state parameter $\omega$. This is done for the inflationary sector of the Supergravity model specified by the potential (\ref{sugrapot4}). As in Fig.~\ref{NreStaro}, the shaded 
region covers the $1-\sigma$ values of $n_s$ according to Planck-2018 \cite{Akrami:2018odb}. In Fig.~\ref{SugraN} we plot, on the contrary, the {\it full} number of $e$-folds $N=N_H+N_{re}$ as a function of $\omega$ for the mean and $1-\sigma$ values of $n_s$ from \cite{Akrami:2018odb}.}
\label{NreSu}
\end{center}
\end{figure}
\begin{figure}[tb]
\begin{center}
\includegraphics[width=8cm]{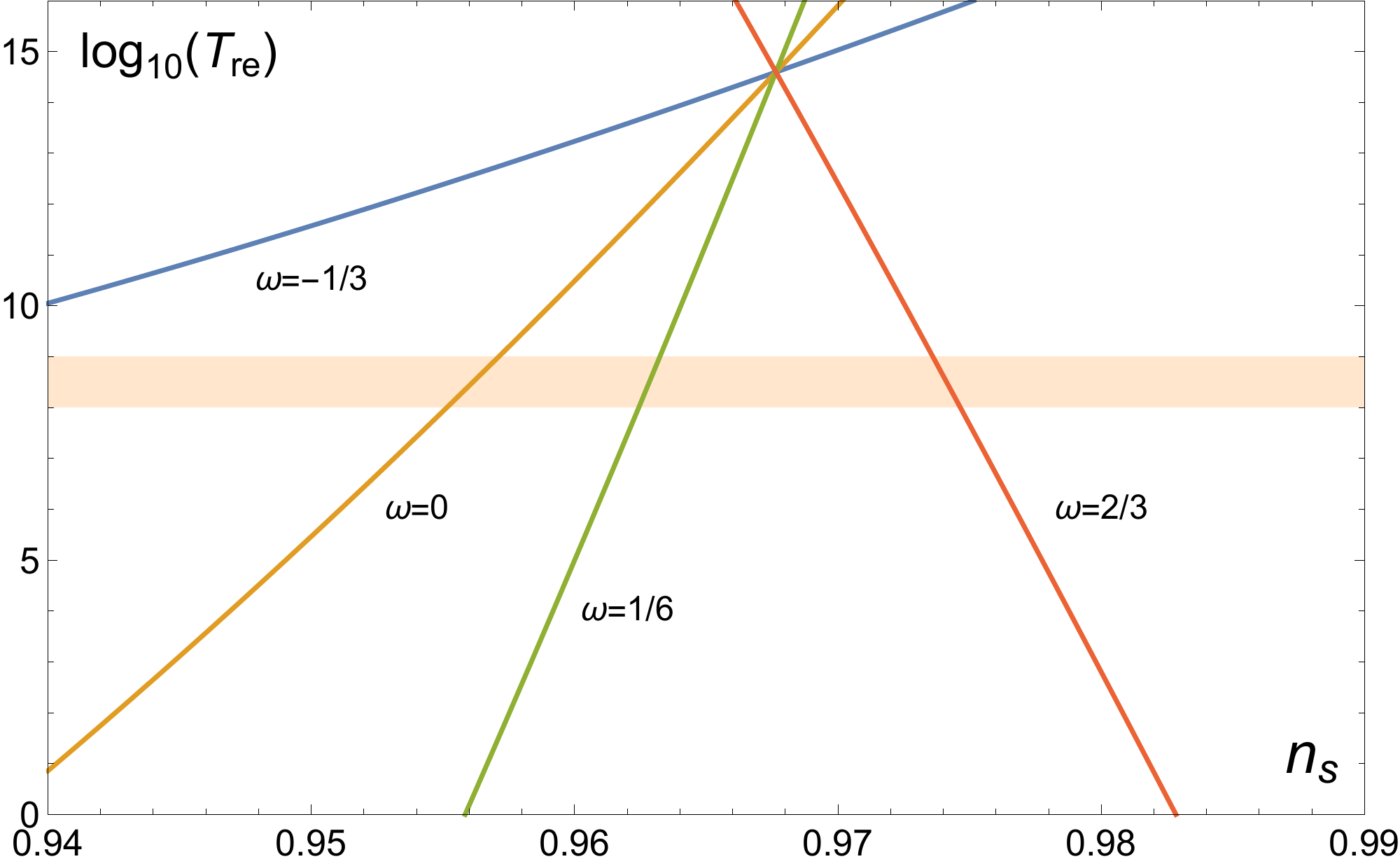}
\caption{\small Plot of the logarithm of the reheat temperature $T_{re}$ for the Supergravity model of Eq.~(\ref{sugrapot4}) as a function of the spectral index $n_s$ as given by Eq.~(\ref{TRE}) for various values of the equation of state parameter $\omega$. The horizontal orange strip indicates the temperature values where PBHs are overproduced if $\omega=0$}\label{TreSu}
\end{center}
\end{figure}
Thus $s$ measures the slope of the potential at the origin. Previous works \cite{Ellis:1982dg,NANOPOULOS198341,OVRUT1983161,HOLMAN1984343,Ross:1995dq} take $s=0$  giving a model presently ruled out by the data \cite{Gomes:2018uhv}. Concretely, the case $s=0$ for $N=60$ $e$-folds implies $n_s \approx 0.934$ which is clearly out of range. On the other hand, imposing $n_s = 0.9649$ implies $N=114$. Also for $s=0$ we get $N_{re}<0$ which is clearly inconsistent. Thus, the case $s=0$ is ruled out and the potential must have some non-vanishing slope {\it at the origin}, as has been previously discussed \cite{German:2004vf}. 

\begin{figure}[tb]
\begin{center}
\includegraphics[width=8cm]{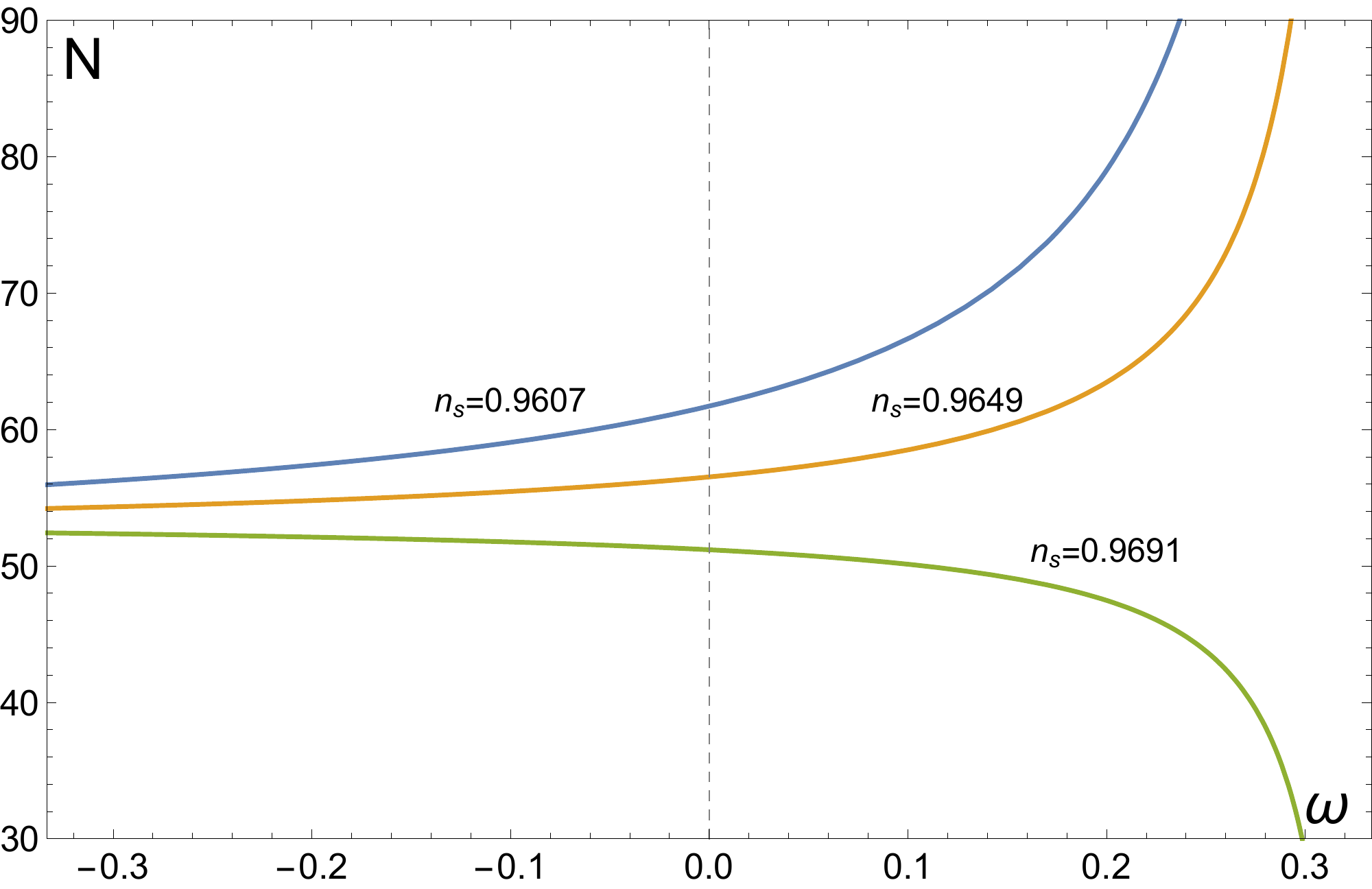}
\caption{\small The case with $s = -1.258\times 10^{-4}$ is here shown for various values of $n_s$. The case $n_s=0.9691$ is already inconsistent for $\omega<1/3$ since a decreasing $N$ implies negative $N_{re}$.}
\label{SugraN}
\end{center}
\end{figure}

\begin{figure}[tb]
\begin{center}
\includegraphics[width=8cm]{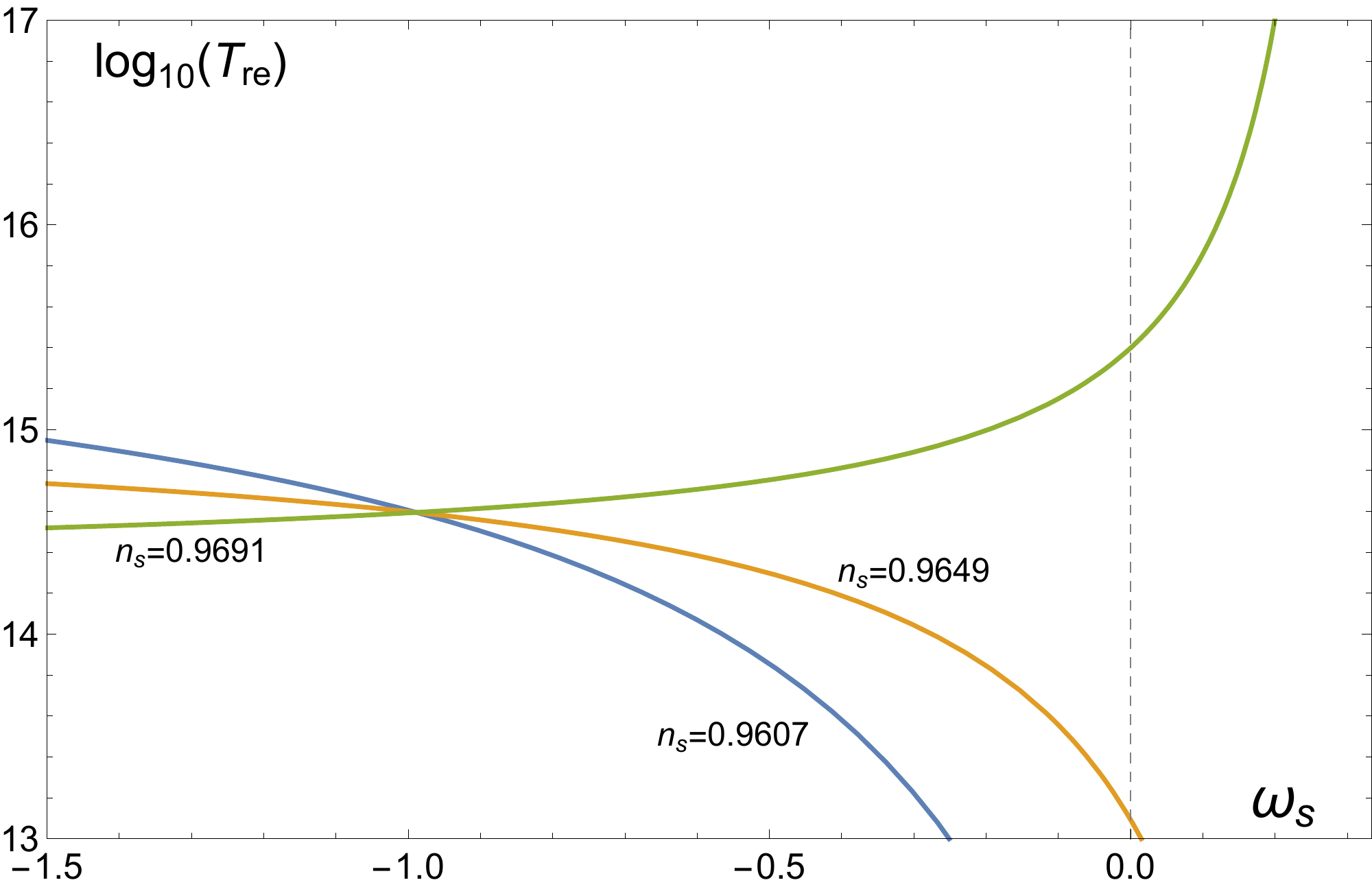}
\caption{\small The case with $s = -1.258\times 10^{-4}$ is here shown for various values of $n_s$. The case $n_s=0.9691$ is already inconsistent for $\omega<1/3$ since  $T_{re}$ should be a decreasing function of $\omega$. }
\label{SugraTre}
\end{center}
\end{figure}

Given that there is no special reason (apart from simplicity) to fix $V^{\prime}(\phi) = 0$ at the origin we allow for an adjustment of parameters and determine the values of $s$ and $\phi_0$ to meet {Planck-2018} requirements. Values of $\phi_0$  very close to $\sqrt{2}$ yield viable realizations of the model \cite{German2018}.
We calculate and plot the number of $e$-folds during reheating and the reheat temperature as functions of $n_s$ in Fig.~\ref{NreSu} and Fig.~\ref{TreSu}, respectively, for the specific case $s=-1.258\times 10^{-4}$. Subsequently, in Fig.~\ref{SugraN} and Fig.~\ref{SugraTre} these same quantities are plotted as functions of the equation of state parameter $\omega$ for values of $n_s$ favoured by Planck-2018. 
For the central value of the spectral index $n_s=0.9649$, the minimum value for the central slope parameter is $s\approx  -1.1852 \times 10^{-4}$; while the set of preferred values ($n_s = 0.9649 \pm 0.0042$) is consistent with $s\geq  -1.2959 \times 10^{-4} $, whenever $\omega<1/3$.
For $s = -1.3348 \times 10^{-4}$, $n_s = 0.9649$ and a {\it full} number of $e$-folds $N=N_H+N_{re}=60$ we get $\omega =0 $ as in the canonical reheating scenario, with the positive number of partial $e$-foldings $N_H=50.78$, $N_{re}=9.22$, The value of the  $T_{re}=4.0 \times 10^{11} GeV$ lies well above of the PBHs constraint. The other observable parameters lie within the bounds imposed by the latest probes as shown in Table~\ref {table2}.
\begin{table}[h!]
  \centering
  \begin{tabular}{ccccccc}
    \\
    $s$ & \quad $\omega_{re}$  &\,\, $N_H$ & \,\,$N_{re}$ & \,\,$r$ & \,\,$n_{sk}$ & \,\, $T_{re}\,(GeV)$
    \\
    \\
    $-1.1868 \times 10^{-4}$ & \quad $0.331$  & \quad $53.02$ & \quad $6.98$ & \quad $1.5 \times 10^{-7}$ & \quad $-2.30 \times 10^{-3}$ & \quad $3.6 \times 10^{11}$\\
   $-1.2959 \times 10^{-4}$ & \quad $0.066$  & \quad $51.33$ & \quad $8.67$ & \quad $1.7 \times 10^{-7}$ & \quad $-2.48 \times 10^{-3}$ & \quad $3.9 \times 10^{11}$\\
  $-1.3348 \times 10^{-4}$ & \quad $0$  & \quad $50.78$ & \quad $9.22$ & \quad $1.8 \times 10^{-7}$ & \quad $-2.55 \times 10^{-3}$ & \quad $4.0 \times 10^{11}$\\
  \\
\end{tabular}
\caption{In all cases the full number of $e$-folds $N\equiv N_H+N_{re}=60$ and $n_s = 0.9649$. By giving values to the slope parameter $s$ we fix $\omega$ and all other quantities follow. For the central value $n_s=0.9649$ the minimum value $s$ can have is $s\approx  -1.1868 \times 10^{-4}$ while the whole range of reported  {\it Planck 2018} values ($n_s = 0.9649 \pm 0.0042$) is consistent with $s\geq  -1.2959 \times 10^{-4}.$ Finally $s = -1.3348 \times 10^{-4}$ gives a canonical reheating escenario $\omega =0$.}
 \label{table2}
\end{table}
\section{Conclusions} \label{Conclusions}

The present paper shows how to constrain inflationary models and reheating by using mixed constraints. On the one hand, we take advantage of the narrow set of spectral index values preferred by the  {Planck-2018} data in order to determine the equation of state parameter $\omega$ during the reheating stage, by fixing the {\it full} number of $e$-folds $N\equiv N_H+N_{re}$ to 60 as required by the inflationary solution to cosmological problems. This is illustrated specifically in Fig.~\ref{N60} for the Starobinsky model and for the most accepted value $n_s = 0.9649$. If the model contains at least one free extra parameter, its value can be set by requiring a canonical reheating scenario $\omega=0$ and imposing $N=60$, with $n_s = 0.9649$. We have exemplified this point with a specific potential of Supergravity inflation presented in Sec.~\ref{Sugra}, with the numerical results given in Table~\ref{table2}. The rest of the observable parameters and the reheating temperature $T_{re}$ follow from the quantities fixed above. We have also shown how the constraints to the PBHs abundance in the case of a dust-like reheating ($\omega \approx 0$) prevent thermalization at a range of temperatures $T_{re}$ relevant for some realisations of the Starobinsky and the Supergravity models (as shown in Figs.~\ref{TreStaro},~\ref{Treheating} and \ref{TreSu}).

In conclusion, the presented method to constrain physical parameters of the reheating and inflationary stages, consisting on matching the observed values of $n_s$ and meeting the requirement of $N>60$ (but not necessarily a fixed value), is suitable to assess the collective viability of inflation plus reheating models. We finally notice that, while we present two explicit examples, our results are robust enough to tackle single-field canonical models of inflation in general.\\

\section*{Acknowledgements}
G.G. acknowledges financial support from UNAM-PAPIIT,  IN104119, {\it Estudios en gravitaci\'on y cosmolog\'ia}. This work is supported in part by SEP-CONACYT grant 282569. The work of A.M. is supported by the postdoctoral grants programme of DGAPA-UNAM.

\bibliographystyle{unsrt}
\bibliography{biblioStaro}

\end{document}